\documentclass{article}
\usepackage{amsfonts,amssymb,amscd}
\usepackage{latexsym, graphicx}
\usepackage[affil-it]{authblk}
\usepackage{algorithmic}
\usepackage{algorithm}

\usepackage{enumerate}

\def\qed{\hfill $\Box$\vspace{2ex}}

\newtheorem{theorem} {Theorem}[section]

\usepackage{xcolor}

\def\inst#1{$^{#1}$}
\begin{document}

\title{Solving the clique cover problem on \newline (bull, $C_4$)-free graphs\thanks{Research
support by Natural Sciences and Engineering Research Council of
Canada.}}
\author{Kathie Cameron\inst{1}
  \and Ch\'inh T. Ho\`ang\inst{2}
}
\date{}
\maketitle
\begin{center}
{\footnotesize

\inst{1} Department of Mathematics, Wilfrid Laurier University,
Waterloo, Ontario,  Canada, N2L 3C5\\
\texttt{kcameron@wlu.ca}

\inst{2} Department of Physics and Computer Science, Wilfrid
Laurier University,
Waterloo, Ontario,  Canada, N2L 3C5\\
\texttt{choang@wlu.ca} }

\end{center}

\begin{abstract}
We give an $O(n^4)$ algorithm to find a minimum clique cover of a (bull, $C_4$)-free graph, or equivalently, a minimum colouring of a (bull, $2K_2$)-free graph, where $n$ is the number of vertices of the graphs.
\end{abstract}

\section{Background}\label{sec:background}
A {\em hole} is a chordless cycle with at least four vertices, and is called odd or even depending on whether
the number of vertices is odd or even. As usual, we will use $C_k$ to denote the hole with $k$ vertices. The {\em bull}
is the graph consisting of a $C_3$ together with two additional vertices of degree 1 adjacent to distinct vertices of the $C_3$.
Where $H$ is a graph, a graph $G$ is called {\em $H$-free} if $G$ has no induced subgraph isomorphic to $H$. 
Where $\mathcal{H}$ is a
set of graphs, a graph $G$ is called {\em $\mathcal{H}$-free} if it is $H$-free for every $H \in \mathcal{H}$. In particular,
(bull, $C_4$)-free graphs are the graphs which have no induced bulls or $C_4$s. 
Where $X$ is a subset of the  vertices of a graph $G$,
we use $G[X]$ to denote the subgraph of $G$ induced by $X$.


Let $G$ be a graph. A {\em clique cover} of $G$ is a set of cliques of $G$ such that every vertex is in at least one of them.
The {\em clique cover number} is the minimum size of a clique cover, and is denoted by $\theta(G)$. The clique cover number
of $G$ equals the chromatic number of its complement, $\overline{G}$. (Note that the minimum size of a clique cover equals the minimum
size of a partition of the vertices into cliques. Somewhat strangely, graph theorists tend to think of partitions into stable sets
and coverings (rather than partitions) by cliques, although in each case, it is clear that 
given a covering one can obtain a partition whose size is
no bigger.) Let $m(G)$ denote the number of edges in a largest
matching of $G$. If $G$ is triangle-free, a minimum clique cover of $G$ consists of a maximum matching together with the
vertices not covered by the matching, and so $\theta(G)=|V(G)|-m(G)$.


A graph is called {\em chordal} if it has no holes.  A vertex is called {\em simplicial} if its neighbours induce a clique.

\begin{theorem}[Dirac \cite{Dir1961}]\label{thm:dirac}
Let $G$ be a chordal graph. Then $G$ is a clique or contains two non-adjacent simplicial vertices.
\end{theorem}

Let $G=(V,E)$ be a graph. An {\it amalgam} partition of $G$ is a partition  of $V$ into disjoint sets $K, A_1, B_1, A_2, B_2$ such that:
\begin{description}
\item[(i)] $A_1 \not= \emptyset$, $ A_2 \not= \emptyset$,
\item[(ii)] $K$ induces a (possibly empty) clique,
\item[(iii)] $| A_i \cup B_i | \geq 2$ for $i=1,2$,
\item[(iv)] there are all possible edges between $A_1$ and $A_2$,
\item[(v)] there are all possible edges between $K$ and $A_i$ for $i = 1,2$,
\item[(vi)] there are no edges between $B_i$ and $A_j \cup B_j$ for $i \not= j$.
\end{description}
We will refer to such a partition by the tuple ($K, A_1, B_1, A_2, B_2$ ).

~

Two vertices are called  {\em true twins} if they are adjacent and they have the same neighbours other than each other. Vertex $x$ is said to {\em dominate} vertex $y$ if every neighbour of $y$ other than $x$ is a neighbour of $x$. We say that $G$ is {\em reducible} if it has adjacent vertices $x$ and $y$ such that $x$ dominates $y$. A {\em universal vertex} is a vertex which is adjacent to all other vertices. Note that a universal vertex dominates all other vertices. Note also that if a graph $G$ is a clique, then every vertex dominates every other vertex. If $v$ is a simplicial vertex and $K$ is its neighbour-set, then any vertex $k \in K$ dominates $v$. Thus, by Dirac's Theorem~\ref{thm:dirac}, a chordal graph is reducible.

A {\em cap} is a hole together with an additional vertex which is adjacent to two adjacent vertices of the hole. Note that a (bull, $C_4$)-free graph is cap-free. A {\em basic} cap-free graph $G$ is either a chordal graph or a biconnected triangle-free graph together with at most one additional vertex, which is adjacent to all other vertices of $G$.

\begin{theorem}[Conforti, Cornu\'ejols, Kapoor, and Vu\v{s}kovi\'c (Thm. 4.1 in \cite{ConCor1999})]\label{thm:cap-free}
A cap-free graph which is not basic contains an amalgam.
\end{theorem}

Note that an irreducible basic graph can not have a universal vertex or be chordal, so it must be a biconnected triangle-free graph.

In this paper, we will show that there is a polynomial-time algorithm to find a minimum clique cover of a (bull, $C_4$)-free graph. Since the bull is self-complementary, this is equivalent to finding a minimum colouring of a (bull, $2K_2$)-free graph, where $2K_2$ is the graph consisting of two independent edges.

Let $P_k$ denote the chordless path on $k$ vertices. The {\em house} is $\overline{P_5}$. The class of (bull, $C_4$)-free graphs is a subclass of (bull, house)-free
graphs, which is a subclass of cap-free graphs.  The complexity of the clique cover problem is unknown for (bull, house)-free graphs and for cap-free graphs.

A {\em k-clique-cover} is a clique cover with at most $k$ cliques (that is, a clique cover of size at most $k$). A $k$-clique-cover (more precisely, a $k$-clique-partition) of a graph corresponds to a $k$-colouring of is complement (that is, a colouring with at most $k$ colours). There is a polynomial-time algorithm for $k$-clique-cover in
(bull, $C_4$)-free graphs \cite{KM}. To explain this, we need some definitions.

An {\em asteroidal triple} or {\em AT} in a graph $G$ is a stable set of three vertices such that between any two, there is a path avoiding the neighbour-set of the third. A graph is {\em AT-free} if it has has no asteroidal triples. As mentioned above, the clique cover problem for (bull, $C_4$)-free graphs is equivalent to the colouring problem for (bull, $2K_2$)-free graphs.  The class of (bull, $2K_2$)-free graphs is a subclass of (bull, $P_5$)-free graphs which are AT-free. Finding a polynomial-time algorithm for colouring AT-free graphs is a long-standing open problem. Stacho \cite{Stacho} gave an $O(n^2m) \le O(n^4)$ algorithm for deciding if an AT-free graph is 3-colourable, and if so, finding a 3-colouring, where $n$ is the number of vertices and $m$ is the number of edges of the input graph. Kratsch and M\"uller \cite{KM} gave an $O(n^{8k+2})$ algorithm for $k$-colouring AT-free graphs. This is polynomial for fixed $k$. Stacho's algorithm solves 3-clique cover and Kratsch and M\"uller's algorithm solves $k$-clique-cover for complements of AT-free graphs, and thus for (bull, $C_4$)-free graphs. Our algorithm can be considered a contribution toward solving the colouring problem for AT-free graphs.

We focus on the clique cover problem for (bull, $C_4$)-free graphs because the complexities of three other fundamental problems - largest clique, stability number and chromatic number - are known. The class of $C_4$-free graphs is known to have a polynomial number of maximal cliques \cite{Farber} and thus the maximum clique and even the maximum weight clique can be found in polynomial time for graphs in this class and thus also for the subclass of (bull, $C_4$)-free graphs. The class of ($C_3,~C_4$)-free graphs is a subclass of (bull, $C_4$)-free graphs. Chromatic number \cite{KraKra2001} and stability number \cite{Poljak}  are NP-hard for ($C_3,~C_4$)-free graphs and thus these problems are also NP-hard for the superclass of (bull, $C_4$)-free graphs.

\section{The results}\label{sec:results}

\begin{theorem}\label{thm:one-point}
Let $G$ be a connected (bull, $C_4$)-free  graph. Then one of the following holds.
\begin{description}
\item [(i)] $G$ is reducible.
\item [(ii)] $G$ is basic.
\item [(iii)] $G$ has a one-point cutset.
\end{description}
\end{theorem}
{\it Proof}. Let $G$ be a (bull, $C_4$)-free  graph. Then $G$ is cap-free. Suppose $G$ is irreducible and not basic. We will show that $G$ has a one-point cutset. By Theorem~\ref{thm:cap-free}, $G$ contains an amalgam $(K, A_1, B_1, A_2, B_2)$. If each $A_i$ contains two non-adjacent vertices (for $i = 1,2$), then there is a $C_4$ in $A_1 \cup A_2$. Thus, w.l.o.g, we may assume $A_2$ is a clique. Suppose that $|A_2| \geq 2$. Then $B_2$ is non-empty, for otherwise, any two vertices in $A_2$ are true twins and so $G$ is reducible, a contradiction. Consider a vertex $a_1 \in A_1$. The vertex $a_1$ must have a neighbour $b_1$ in $B_1$, for otherwise, any vertex $a_2 \in A_2$ dominates $a_1$, a contradiction. Consider two vertices $x,y \in A_2$. Since $G$ is irreducible, there must be a vertex $z$ that is adjacent to $x$ but not to $y$. The vertex $z$ must lie in $B_2$. But now $G[ \{b_1, a_1, x,y,z \}]$ is a bull, a contradiction. So we have $|A_2| = 1$ and therefore $|B_2| \geq 1$.

Suppose that $K \not= \emptyset$. Consider  vertices $k \in K$ and $x \in A_2$. Since $G$ is irreducible, there must be a vertex $z$ that is adjacent to $x$ but not to $k$. The vertex $z$ is necessarily in $B_2$. A vertex $a_1 \in A_1$ must have a neighbour $b$ that is not adjacent to $k$ since $G$ is irreducible. The vertex $b$ must lie in $B_1$. But now $G[\{ z,x,k,a_1,b\}]$ is a bull. So, $K$ is an empty set. Now, $A_2$ is a one-point cutset of $G$. \qed
\begin{theorem}\label{thm:stable}
Let $G$ be a connected, irreducible (bull, $C_4$)-free graph with a one-point cutset $v$. Then the neighbourhood of $v$ is a stable set.
\end{theorem}
{\it Proof}. Enumerate the components of $G-v$ as $C_1, C_2, \ldots, C_t$. Define $N_i = N(v) \cap C_i$. We only need to prove that $N_i$ is a stable set. Note that each vertex $x \in N_i$ must have a neighbour in $M_i = C_i - N_i$ since otherwise $v$ dominates $x$. In particular, $M_i \not= \emptyset$. Let us assume that $N_i$ is not a stable set.  Consider a component $C$ of $G[N_i]$ with at least two vertices.

\noindent
Claim: For any two adjacent vertices $x,y \in C$, $N(x) \cap M_i = N(y) \cap M_i$. \\
{\it Proof}. If the claim is false then there is a vertex $z \in M_i$ that is adjacent to $x$ but not to $y$  (or, vice versa). But now $G[\{z,x,y,v,c_j\}]$ is a bull for a neighbour $c_j$ of $v$ in $C_j$ with $j \not = i$. The claim is justified.

Now, $C$ must be a clique, for otherwise, consider two non-adjacent vertices $x,y \in C$. There is a path $P$ in $C$ joining $x$ and $y$. By the claim, adjacent vertices of  $P$ have the same neighbours in $M_i$. It follows that all vertices of $P$ have the same neighbours in $M_i$. As mentioned above, there is always at least one such neighbour.  Thus $x$ and $y$ have a common neighbour, say $z$ in $M_i$, and then $G[\{v,x,y,z\}]$ is a $C_4$, a contradiction. Since $C$ is a clique, any two vertices of $C$ form a pair of true twins, a contradiction. Thus each component $C$ of $G[N_i]$ is a single vertex and so $N_i$ is a stable set. \qed

Let $v$ be a one-point cutset of $G$. Let the components of $G - v$ be $C_1, C_2 , \ldots, C_t$. Define $f(v) =$ minimum $\{|C_i|$ : $i = 1, 2, \ldots t\}$.  We say that $v$ is {\it terminal} if there is a component $C_i$ such that $C_i \cup \{v\}$ induces a triangle-free basic graph. We now strengthen Theorem \ref{thm:one-point} to the following:

\begin{theorem}\label{thm:terminal-one-point}
Let $G$ be a connected (bull, $C_4$)-free graph. Then one of the following holds .
\begin{description}
\item [(i)] $G$ is reducible.
\item [(ii)] $G$ is basic.
\item [(iii)] $G$ has a terminal one-point cutset.
\end{description}
\end{theorem}
{\it Proof}.
Let $G$ be a connected (bull, $C_4$)-free graph. Suppose $G$ is irreducible and not basic. By Theorem~\ref{thm:one-point}, $G$ has a one-point cutset. Among all one-point cutsets of $G$, choose the one, $v$, with the smallest $f(v)$.
Let $C_i$ be a component of $G-v$ with $f(v) = | C_i |$. Let $G_i$ be the subgraph of $G$ induced by $C_i \cup \{v\}$.  By Theorem~\ref{thm:stable}, the neighbourhood $N_i$ of $v$ in $G_i$ is a stable set. Every vertex $u \in N_i$ must have a neighbour in $M_i = C_i - (N_i \cup \{v\}) $, for otherwise $v$ dominates $u$ in $G$, a contradiction.
We have $| N_i | \geq 2$, for otherwise,
the unique neighbour $v'$ of $v$ is a one-point cutset of $G_i$, and it is also a one-point cutset of $G$ with $f(v') < f(v)$, a contradiction to the minimality of $f(v)$. By Theorem~\ref{thm:one-point}, $G_i$ is reducible, basic, or has an one-point cutset.

Suppose $G_i$ has a one-point cutset $v_i$. Then, $v_i$ is also a one-point cutset of $G$ with $f(v_i) < f(v)$, a contradiction to the minimality of $v$. So $G_i$ cannot have a one-point cutset.

Suppose $G_i$ is reducible and has no one-point cutset. Let $x$ and $y$ be two adjacent vertices in $G_i$ such that $x$ dominates $y$.   Since $N_i$ is a stable set of size at least two, no vertex in $N_i$ can dominate $v$, that is, $y \not = v$.
If $y \in N_i$, then $x$ is necessarily $v$, and so $x$ dominates $y$ in $G$, a contradiction. Thus, $y$ is in $M_i$ and so it has no neighbours in $G-G_i$. But now $x$ dominates $y$ in $G$, a contradiction.

Thus $G_i$ is basic, irreducible and has no one-point cutset. Since $G_i$ is irreducible and basic, it must be triangle-free. Thus $v$ is a terminal one-point cutset.\qed

Remark. We can find in polynomial-time a terminal one-point cutset if one exists in $O(nm)$ time, where, as usual, $n$, repectively, $m$,  is the number of vertices, respectively, edges, of the input graph.

Remark. For a triangle-free graph $G$, there is a polynomial-time algorithm to find $\theta(G)$. (Find a maximum matching.)

\begin{theorem}\label{thm:clique-cover}
There is a polynomial-time algorithm to find a minimum clique cover for a (bull, $C_4$)-free graph.
\end{theorem}
{\it Proof}. Let $G$ be a (bull, $C_4$)-free graph. We may assume that $G$ is connected. If $G$ contains adjacent vertices $x$ and $y$ such that $x$ dominates $y$ then we remove $x$ from $G$, and it is easy to see that $\theta(G) = \theta(G-x)$. We repeatedly remove such vertices $x$ until $G$ becomes irreducible. If $G$ is basic then it is triangle-free, and we can compute $\theta(G)$ directly.

Now, suppose $G$ has a terminal one-point cutset $v$. Let $C_i$ be a component of $G-v$ such that the subgraph $G_i$ of $G$ induced by $C_i \cup \{v\}$ is triangle-free. Let $G'$ be the subgraph of $G$ induced by the vertices in $G- C_i$.
Compute $m(G_i)$ and $m(G_i - v)$. If $m(G_i) = m(G_i-v)$, then there is a minimum clique cover ${\cal C}$ of $G_i$ such that $\{v\}$ is a member of ${\cal C}$. We have thus $\theta(G) = \theta(G') + \theta(G_i - v)$. We can recursively compute $\theta(G')$ to determine $\theta(G)$. Now, we may assume $m(G_i) = m(G_i - v) + 1$. This means for every maximum matching ${\cal M}$ of $G_i$, some edge of ${\cal M}$ is incident to $v$. In other words, for any minimum clique cover ${\cal C}$ of $G_i$, the vertex $v$ belongs to a clique (of ${\cal C}$) of size 2. Thus we have $\theta(G_i) = \theta(G_i - v)$. It follows that $\theta(G) = \theta(G'-v) + \theta(G_i)$. We can recursively compute $\theta(G'-v)$ to determine $\theta(G)$. Since the recursion is done at most $n$ times, the algorithm is polynomial. \qed

\noindent
\textbf{Complexity of the algorithm.} \\
We can check if $G$ is connected in $O(n+m)$ time, and if not, apply the algorithm to each component of $G$.\\
For each vertex $x$ and for each neighbor $y$ of $x$, we can check in $O(m)$ time if $x$ dominates $y$, or vice versa. 
Thus, we  can find a dominating vertex, if one exists, in $O(nm)$ time.  \\
We can check if the remaining graph $H$ is triangle-free in $O(n^\alpha)$ time, 
where $\alpha$ is the complexity of matrix multiplication 
which is currently 2.3728639 \cite{LeG2014}.  If $H$ is triangle-free, then we can find a minimum clique cover of $H$ using the matching algorithm.\\
Since the graph $H$ is irreducible, if it is not triangle-tree, then it is not basic, so by Theorem \ref{thm:terminal-one-point}, it contains a terminal one-point cutset $v$. As mentioned above, a terminal one-point cutset can be found in $O(nm)$ time (by checking for each vertex $v$, whether it is a one-point cutset, and if so, computing the function $f(v)$ and choosing $v$ with the smallest value $f(v)$).
%
Where $G_i$ is the subgraph of $H$ induced by $C_i \cup \{v\}$, compute $m(G_i)$ and $m(G_i - v )$. This can be done in $O(\sqrt{n}m)$ time \cite{matching}.\\
Now the algorithm is iterated, at most $n$ times.\\
The overall complexity is:
$n[O(n+m)+ \; O(nm)+$ $O(n^\alpha)+$ $O(\sqrt{n}m)+O(nm)+ \; O(\sqrt{n}m)]$ $= O(n^2 m + n^{\alpha + 1}) = O(n^4)$.

\section{Open problems}\label{sec:problems}
As mentioned in Section \ref{sec:background}, the complexity of the clique cover problem is unknown for (bull, house)-free graphs and for two different superclasses of these: cap-free graphs and complements of AT-free graphs.

\end{document}